\documentclass[twocolumn,prl,a4paper,showpacs,floatfix,preprintnumbers,amsmath,amssymb]{revtex4}

\usepackage{amsmath}
\usepackage{graphicx}

\usepackage{graphicx}
\usepackage{dcolumn}
\usepackage{bm}

\newcommand{\ud}{\mathrm{d}}

\newcommand{\be}{\begin{equation}}
\newcommand{\ee}{\end{equation}}

\begin{document}


\title{Mach-Zehnder Interferometry 
at the Heisenberg Limit\\ 
with coherent and squeezed-vacuum light}

\author{Luca Pezz\'e and Augusto Smerzi}
\affiliation{ BEC-CNR-INFM and Dipartimento di Fisica, Universit\`a di Trento, I-38050 Povo, Italy}

\date{\today}
                      
\begin{abstract}
We show that the phase sensitivity $\Delta \theta$ of a Mach-Zehnder interferometer 
fed by a coherent state in one input port and squeezed-vacuum in the other one is 
i) independent from the true value of the phase shift and ii) can reach the 
Heisenberg limit $\Delta \theta \sim 1/N_T$, where $N_T$ is the average number of particles of the input states.
We also show that the Cramer-Rao lower bound,  $\Delta \theta \propto 1/ \sqrt{|\alpha|^2 e^{2r} + \sinh^2r}$,
can be saturated for arbitrary values of the squeezing parameter $r$ and the amplitude of the coherent mode $|\alpha|$ by a Bayesian phase inference protocol. 
\end{abstract}

\maketitle

\begin{figure}[!t]
\begin{center}
\includegraphics[scale=0.45]{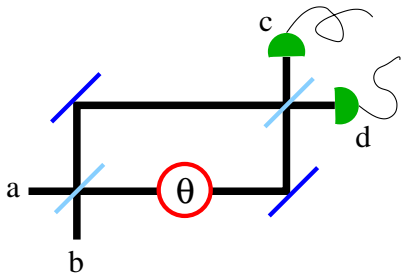}
\end{center}
\caption{\small{(Color online) Schematic representation of the Mach-Zehnder interferometer.
The input modes $a,b$ are a coherent and a squeezed-vacuum field, respectively.}} \label{mz} 
\end{figure}

{\it Introduction}. 
The goal of quantum interferometry is to estimate phases
beyond the shot-noise (``standard quantum") limit. 
The quest requires proper non-classical states,
as was first shown by Caves in 1981 \cite{Caves_1981}, who considered a Mach-Zehnder (MZ)
fed by coherent $\otimes$ squeezed-vacuum light.
This benchmark generated a large body of theoretical  
\cite{Bondurant_1984, Gea_1987, Paris_1995, scully} and 
experimental \cite{Xiao_1987, Wu_1986, Breitenbach_1997} studies,
including the demonstration of sub shot-noise sensitivity \cite{Xiao_1987} 
using parametric down-conversion in a cavity as a source of squeezed vacuum \cite{Wu_1986}.
The scheme proposed by Caves is sketched in Fig.(\ref{mz}). 
One of the inputs of the linear loss-less MZ is the coherent state
$|\alpha\rangle_a \equiv \sum_{m=0}^{+\infty} C_m |m\rangle_a$, with $\alpha \equiv e^{i\theta_c}|\alpha|$
and $C_m \equiv \frac{\alpha^m e^{-|\alpha|^2/2}}{\sqrt{m!}}$. 
The second input state is the squeezed-vacuum
$| \zeta \rangle_b \equiv \sum_{m=0}^{+\infty}S_{m}|m\rangle_b$, with $\zeta \equiv r e^{i\theta_s}$ and
$S_m\equiv \frac{(e^{i \theta_s} \tanh r )^{m/2}}{2^{m/2}\sqrt{ m !\cosh r}}\mathrm{H}_{m}(0)$ \cite{Yuen_1976},
$\mathrm{H}_{m}(x)$ being the Hermite polynomials. 
Following the current literature, based on the original works of the 80's,
the phase estimate of this system is retrieved from the measurement of the relative number of particles at the 
output ports $\hat{M}_{out}=\hat{N}_c-\hat{N}_d$.
Fluctuations on the results obtained in $p$ independent measurements 
propagate to the estimated value of the
phase shift $\theta$ \cite{nota1}, which can be eventually determined with uncertainty \cite{nota2}:  
\begin{eqnarray} \label{SQ.1}
\Delta \theta = \frac{1}{\sqrt{p}} \sqrt{\frac{|\alpha|^2 e^{-2r} + \sinh^2 r}{\big( |\alpha|^2 - \sinh^2 r \big)^2} +
\frac{|\alpha|^2 + 2 \sinh^2 r \cosh^2 r}{\big( |\alpha|^2 - \sinh^2 r \big)^2 \tan^2 \theta }}. 
\end{eqnarray}
According to Eq.(\ref{SQ.1}), we can appreciate an increase
of phase sensitivity with respect to the shot-noise
only when the true value of the phase shift is sufficiently close to 
$\theta = \pi/2$ \cite{Caves_1981, scully} (dark fringe), where 
$\langle \hat{M}_{out} \rangle =\langle \hat{N}_c-\hat{N}_d \rangle= 0$. 
On the other hand, $\langle \hat{M}_{out} \rangle$ 
depends weakly on the phase shift when $\theta \approx 0, \pi$ and 
the error propagation formula Eq.(\ref{SQ.1})
predicts large phase fluctuations around these points. 
Asymptotically in the amplitude of the coherent state,  $|\alpha|^2 \gg \sinh^2 r$, and for a fixed
squeezing parameter $r$, 
Eq.(\ref{SQ.1}) predicts a sub shot-noise sensitivity \cite{Caves_1981, nota3}  
\be \label{SQ.3}
\Delta \theta = \frac{1}{\sqrt{p}} \frac{e^{-r}}{\sqrt{\bar{n}}}, \qquad\qquad (\theta=\pi/2),
\ee 
with the average number of photons injected in the MZ $\bar{n} \simeq |\alpha|^2$. 

In this Letter we show that the choice of the average relative number of 
photons as phase estimator is not optimal. Quantum fluctuations also contains information on
the true value of the phase shift, which can be retrieved by taking in account the higher 
moments of the measured number of particles at the output ports. 
We will show that the ultimate phase sensitivity of a Mach-Zehnder fed by  
coherent $\otimes$ squeezed-vacuum light is
\be \label{SQ.4}
\Delta \theta = \frac{1}{\sqrt{p}} \frac{1}{ \sqrt{|\alpha|^2 e^{2r} + \sinh^2r}} \qquad (0\leq\theta\leq\pi).
\ee
The phase sensitivity Eq.(\ref{SQ.4}) is i) independent from the true value of the phase shift over 
the whole interval $0\leq\theta\leq\pi$ and ii) it reaches, at the optimal point $|\alpha|^2= \sinh^2r$, the Heisenberg limit:
\be \label{HL}
\Delta \theta = \frac{1}{\sqrt{p}} \frac{1}{\bar{n}}, \qquad\qquad (0\leq\theta\leq\pi),
\ee
asymptotically in the average number of photons $\bar{n} = |\alpha|^2 + \sinh^2r$ 
and with a number of independent measurements $p \gtrsim 30$.

In the following, we will first analytically calculate the Cramer-Rao lower bound (CRLB), 
Eq.(\ref{SQ.4}), and then demonstrate that it is saturated by a Bayesian phase inference approach.   
A proof of principle of Eq.(\ref{HL}) can be obtained within current technology, 
at least in the limit of small $\bar{n}$: 
high-efficiency number-resolving photo-detectors have been recently 
applied to interferometry \cite{Pezze_2007,Khoury_2006}
and high squeezing has been obtained with parametric down-conversion \cite{Eisenberg_2004}.
Out results can be relevant, for instance, to improve the
efficiency of the large scale interferometers dedicated to the detection of gravitational waves
\cite{Hough_2005}, which would not require phase-stabilization techniques  to lock at the optimal point
$\theta=\pi/2$ \cite{Gea_1987} and which can significantly increase their sensitivity.

\begin{figure}[!t]
\begin{center}
\includegraphics[scale=0.52]{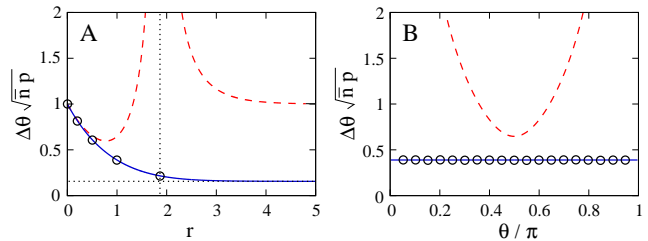}
\end{center}
\caption{\small{(Color online) 
Comparison between Eq.(\ref{SQ.1}) (dashed red line) and Eq.(\ref{SQ.4}) (solid blue line).
Circles are the results of the Bayesian analysis.
A) Phase sensitivity $\Delta \theta \sqrt{\bar{n}p}$ as function of the squeezing parameter $r$, for $\theta=\pi/2$
and $|\alpha|^2=10$.
Notice that Eq.(\ref{SQ.1}) diverges at $\sinh^2 r = |\alpha|^2$ (dotted vertical line).  
For $r\gg 1$, Eq.(\ref{SQ.4}) gives $\Delta \theta \sqrt{\bar{n}p} \to 1/\sqrt{4|\alpha|^2+1}$ (dotted horizontal line).
B) $\Delta \theta \sqrt{\bar{n}p}$ in the limit $p \to \infty$ as a function of the true value of the phase shift.
Here $|\alpha|^2=10$ and $r=1$.}}
\label{figure1} 
\end{figure}

{\it The Cramer-Rao lower bound}. 
The output state of a loss-less Mach-Zehnder interferometer 
is given by $|\psi_{out}\rangle=e^{-i \theta \hat{J}_y}|\psi_{in}\rangle$ \cite{nota4}, 
where, in our case, $|\psi_{in}\rangle=|\alpha\rangle_a|\zeta\rangle_b$.
The conditional probability to measure $N_c$ and $N_d$ particles at the output ports, 
given an unknown phase shift $\theta$, is 
\be \label{Prob}
P(N_c,N_d|\theta) = \bigg| \sum_{n=0}^{N} 
C_{N-n} \,
S_n \, 
d_{\mu,N/2-n}^{N/2}(\cos \theta) 
\bigg|^2,
\ee
where $\mu=(N_c-N_d)/2$ and $d^j_{\mu, \nu}(\cos \theta)$ are rotation matrix elements.  
The Fisher information, $\mathrm{F}(\theta) = \sum_{N_c=0}^{+\infty} \sum_{N_d=0}^{+\infty}
\frac{1}{P(N_c,N_d|\theta)} \big( \frac{ \partial P(N_c,N_d|\theta)}{ \partial \theta} \big)^2$,
turns out to be independent from the true value of the phase shift $\theta$, see Fig.(\ref{figure1}),
and an analytical calculation gives $\mathrm{F}(\theta)=|\alpha|^2 e^{2r} + \sinh^2r$.  
According to Cramer and Rao, the phase sensitivity of an unbiased estimator is bounded by 
$\Delta \theta = \frac{1}{\sqrt{p\mathrm{F}(\theta)}}$, which, after replacing the previous expression 
for the Fisher information, gives Eq.(\ref{SQ.4}). 
There are interesting limit regimes recovered by this equation: 
i) When $r=0$ {\it or} $\alpha=0$ we get the ($\theta$-independent) shot-noise limit $\Delta \theta = 1/\sqrt{p\bar{n}}$. 
The phase independence of the case $r=0$ has been studied and experimentally demonstrated in \cite{Pezze_2007}.
ii) When $\sinh^2 r \ll |\alpha|^2$ we obtain the sub shot-noise limit discussed by Caves, 
$\Delta \theta = e^{-r}/\sqrt{p\bar{n}}$ Eq.(\ref{SQ.3}) with, again, the important 
difference that, here, the phase sensitivity is independent of the phase shift for
$0\leqslant \theta \leqslant \pi$.
Notice that, in the limit of very high squeezing, $\sinh^2r \gg |\alpha|^2$, Eq.(\ref{SQ.1}) predicts  
$\Delta \theta = 1/\sqrt{p\bar{n}}$ (still at $\theta=\pi/2$), while Eq.(\ref{SQ.4}) gives a sub shot-noise 
scaling $\Delta \theta = 1/(\sqrt{p\bar{n}}\sqrt{4|\alpha|^2+1})$.

The most important regime predicted by Eq.(\ref{SQ.4}) is obtained when $|\alpha|^2 \sim \sinh^2 r = \bar{n}/2$ 
(i.e., with half of the input intensity provided by the coherent state and half by the squeezed light). This gives
$\Delta \theta = 1/\sqrt{p} \bar{n}$ when $\bar{n},p \gg 1$.
It is interesting to notice that, for these optimal values of the parameters $\alpha$ and $r$, 
the error propagation formula Eq.(\ref{SQ.1}) predicts a divergence. 
In figure (\ref{figure1},a) we compare, as a function of $r$ and for $\theta  = \pi/2$,
the quantity $\sqrt{\bar{n}p}~\Delta \theta$ calculated with Eq.(\ref{SQ.1})
(dotted line) compared with Eq.(\ref{SQ.4}) (solid line). 

Why does the error propagation formula Eq.(\ref{SQ.1}) provides such a poor phase sensitivity with respect to 
the CRLB?
The answer is that an estimation of the phase shift based only on the measurement of the average relative 
number of particles does not exploit all the available information contained in the detection of 
$N_c$ and $N_d$. It forgets about the information contained in the fluctuations of the total number of particles and
(since the relevant probability distributions are not Gaussians) in the higher moments \cite{nota5}.  
In figure (\ref{figure1},b) we plot the phase sensitivity $\sqrt{\bar{n}p}~\Delta \theta$ as a function 
of the true value of the phase shift. 
The dashed line is the result Eq.(\ref{SQ.1}) and the solid line is Eq.(\ref{SQ.4}).

\textit{Bayesian analysis.} Is it possible to saturate the CRLB and demonstrate 
a phase sensitivity at the Heisenberg limit $\Delta \theta \sim 1/N_T$, being $N_T$ the average number of 
particles burnt during the estimation process \cite{nota6}?
A possibility, of course, is to consider the Maximum Likelihood estimator which, according to the 
Fisher theorem, saturates the CRLB asymptotically in the number of measurements $p$.
In the following, however, we consider a Bayesian protocol \cite{Pezze_2006} showing that it also saturates the CRLB.
To simulate a phase estimation experiment, we i) randomly choose $p$ values $N_c^{(i)}, N_d^{(i)}$ 
at the output ports distributed according to $P(N_c,N_d|\theta)$ with an unknown $\theta$;
ii) invert, by applying the Bayes theorem, the distribution Eq.(\ref{Prob}) and associate 
to the measured values $N_c^{(i)}, N_d^{(i)}$  the probability distribution 
$P(\phi|{N_c^{(1)}, N_d^{(1)};...;N_c^{(p)}, N_d^{(p)}}) \sim \prod_{i=1}^{p}P(\phi|N_c^{(i)},N_d^{(i)})$; 
iii) calculate the phase sensitivity as 68\% confidence around the maximum of the phase distribution.
In Fig.(\ref{figure1},a,b), the circles, obtained with the Bayesian probabilities 
asymptotically in the number of independent measurements $p$, coincide with 
the analytical expression of the CRLB, Eq.(\ref{SQ.4}). 

\begin{figure}[!t]
\begin{center}
\includegraphics[scale=0.53]{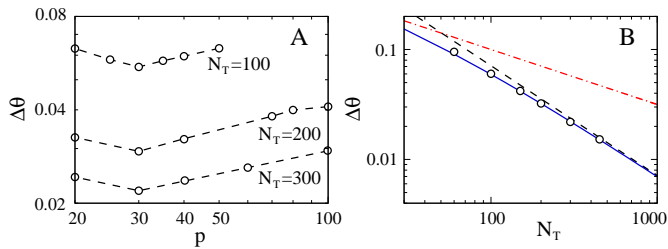}
\end{center}
\caption{\small{(Color online) Demonstration of the Heisenberg limit $\Delta \theta \sim 1/N_T$.
A) Circles are the phase sensitivity obtained with the Bayesian analysis 
as a function of the number of measurements $p$
with fixed total number of particles $N_T=p\bar{n}$.
The optimal value, $p_{opt}=30$, corresponds to the minimum of $\Delta \theta$ and
does not depend on $\bar{n}$. Dashed lines are guides to the eye. 
B) Corresponding optimal sensitivity as a function of $N_T$. 
The dashed line is the asymptotic limit $\Delta \theta = 7.12/N_T$,  
the solid line is $\Delta \theta \sim \frac{1}{\sqrt{p_{opt}} \sqrt{|\alpha|^2 + \sinh^2 r} }$.
Shot-noise has been included for comparison (dot-dashed line).}}
\label{figure6} 
\end{figure}

Yet, in order to demonstrate the possibility to reach the Heisenberg limit, $\Delta \theta \sim 1/N_T$, 
we have to carefully analyze the role of $p$ \cite{Braunstein_1992}.
Within the optimal choice of parameters, $|\alpha|^2 \sim \sinh^2 r \sim \bar{n}/2 \gg 1$,
we fix a total number of particles, $N_T = p \bar{n}$, distributed in ensembles of $p$ independent measurements. 
There are two concurring behaviors contributing, in average, to $\Delta \theta$: 
for small $p$ (large $\bar{n}$) we are in a pre-asymptotic regime
characterized by large oscillations of $\sum_{i=1}^p N_c^{(i)}+ N_d^{(i)}$, which still provides 
sub shot-noise but not the Heisenberg limit. 
For larger values of $p$, we saturate the Fisher information and obtain 
$\Delta \theta = \sqrt{p}/N_T$. 
The prefactor $\sqrt{p}$ arises from the statistics of independent measurements.   
As shown in figure (\ref{figure6},a), the optimal value is $p_{opt}\sim 30$. 
The crucial point to notice is that $p_{opt}$ does not depend on $N_T$. If it would, we could not
claim the Heisenberg limit.
The phase sensitivity calculated at  $p_{opt}$ is plotted in figure (\ref{figure6},b) as a function of 
$N_T$ (circles). The dashed line is the Heisenberg limit $\Delta \theta = 7.12/N_T$, while the 
solid line is $\Delta \theta \approx \frac{1}{\sqrt{p_{opt}} \sqrt{|\alpha|^2 + \sinh^2 r} }$. 
For comparison, we include in the figure the shot-noise limit (dot-dashed line).

We emphasize that an enhancement of phase sensitivity 
can be obtained also when only one output port is monitored 
(reduced MZ configuration). 
A numerical calculation of the Fisher information for $|\alpha|^2\sim \sinh^2 r$
shows a strong dependence on $\theta$, the optimal working point being close to $0$ or $\pi$, 
depending on the port which is monitored. 
Even if we were not able to numerically investigate large values 
of $\bar{n}$, we have strong evidences that, asymptotically in $\bar{n}$, we obtain 
a phase sensitivity $\Delta \theta \sim 1/N_T$, with a prefactor larger than the one 
obtained with the Mach-Zehnder interferometer.

\begin{figure}[!t]
\begin{center}
\includegraphics[scale=0.6]{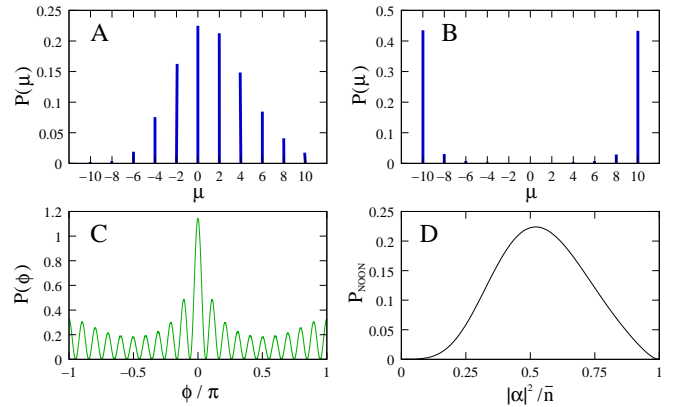}
\end{center}
\caption{\small{ (Color online)
Relative number of particles distribution $P(\mu)$ for A) the input state $|\psi_N\rangle$ 
with post-selected $N=\bar{n}$ and optimal conditions $|\alpha|^2=\sinh^2r$ and 
B) state after the first beam splitter, $|\psi_N^{BS}\rangle$.
C) Phase distribution $P(\phi)$ obtained after a projection of $|\psi_N^{BS}\rangle$ over phase states. 
D) Quantity $P_{NOON}$ as a function of $|\alpha|^2/\bar{n}$. The maximum is reached at $|\alpha|^2\sim\bar{n}/2$.
Here $\bar{n}=20$.}}\label{figure5} 
\end{figure}

{\it Discussion.}
What is the physics underlying the increase in phase sensitivity using squeezed vacuum light?
In \cite{Caves_1981} Caves associated sub shot-noise to quadrature squeezing. 
Indeed, under the conditions $\theta=\pi/2$, 
and $|\alpha|^2\gg \sinh^2 r$, 
Eq.(\ref{SQ.1}) reduces to $\Delta \theta = \frac{\Delta \hat{X}_1}{\sqrt{p \bar{n}}}$, with the 
quadrature $\hat{X}_1=(\hat{b}^{\dag}+\hat{b})/2$.
With squeezed-vacuum light $\Delta \hat{X}_1=e^{-r}$ and we recover Eq.(\ref{SQ.3}).
Conversely, we can understand the saturation at the Heisenberg limit by quantum
interference effects created by the beam splitter. The key point is to notice that 
the input squeezed state has the components 
$S_m=0$ when $m$ is odd.
For the sake of simplicity, we discuss this problem by fixing (post-selecting) a total number of particles $N = \bar{n}$.
The input $|\psi_N\rangle\equiv \sum_{\mu=-N/2}^{N/2}A_{\mu} |N/2-\mu\rangle_a|N/2+\mu\rangle_b$ 
is characterized by a relative number of particles distribution 
$P(\mu)=|A_{\mu}|^2$, where $A_{\mu}=0$ for odd values of $N/2-\mu$, see Fig.(\ref{figure5},a). 
This creates a relative number of particles distribution after the first beam splitter
characterized by a mean-square fluctuation of the order of $N$. In particular, the distribution has
the largest peaks centered at $\mu = \pm N/2$, see Fig.(\ref{figure5},b), which indicates that 
the corresponding quantum state after the beam splitter contains a large 
``NOON" component $|NOON\rangle \sim (|N,0\rangle+|0,N\rangle)$.
Such a distribution is typical of states attaining the Heisenberg limit $\Delta \theta \sim 1/N$.
Intuitively, the phase distribution, $P(\phi)$ obtained by projecting a state with 
heavily weighted components at $\mu = \pm N/2$ over 
phase states $|\phi\rangle=\sum_{\nu=-N/2}^{N/2}e^{-i \nu \phi} |N/2- \nu\rangle_a|N/2+\nu\rangle_b$,  
is characterized by oscillations of frequency $2\pi/N$.
This typical structure is illustrated in Fig.(\ref{figure5},c) where 
we plot the phase distribution obtained by projecting 
$|\psi_N^{BS}\rangle = e^{-i\frac{\pi}{2}\hat{J}_x}|\psi_N\rangle$ over $|\phi\rangle$.
Finally, it is interesting to notice that the highest ``NOON" component is obtained
when $\alpha^2=\bar{n}/2$, which precisely corresponds to the optimal 
conditions discussed in Eq.(\ref{HL}). This is illustrated in Fig.(\ref{figure5},d), 
where $P_{NOON}\equiv |\langle NOON | e^{-i\frac{\pi}{2}\hat{J}_x}|\psi_N\rangle|^2$ is shown
as a function of $|\alpha|^2/\bar{n}$.

{\it Conlusions.}
The discovery that interferometric measurements can be dramatically improved by non-classical light 
has been crucial for the development of modern quantum optics \cite{scully, Dowling_1998}. 
Several states and strategy have been proposed in the literature to beat the shot-noise limit.
Here we have shown that the oldest of these proposals, a linear lossless Mach-Zehnder interferometer fed by 
a coherent$\otimes$squeezed-vacuum light \cite{Caves_1981}, can indeed reach the Heisenberg limit Eq.(\ref{HL}),
but \emph{only if} the whole information included in the measurement of the 
number of particles at the output ports is taken into account. 
This requires a feasible analysis of the interferometric data which is provided, for instance, 
by a Bayesian protocol.
Moreover, we have also shown that the phase sensitivity is independent from the true 
value of the phase shift for arbitrary values of squeezing.


\begin{thebibliography}{99}
\bibitem{nota1} Being $A_p=\sum_{i=1}^p (N_c^{(i)}-N_d^{(i)})/p$ the result 
obtained after $p$ independent measurements, 
we can write $A_p=\langle \hat{M}_{out} \rangle \pm \frac{\Delta \hat{M}_{out}}{\sqrt{p}}$, 
where $\langle \hat{M}_{out}\rangle= \langle \hat{N}_c-\hat{N}_d\rangle$ and 
$(\Delta \hat{M}_{out} )^2$ are the average value and the variance of the relative number of particles 
at the output ports, respectively. 
Fluctuations of $A_p$ in different independent measurements 
done under the same initial conditions propagates to the estimated phase shift. 
A linear error propagation gives 
$A_p=\langle \hat{M}_{out} \rangle \pm \frac{\partial \langle \hat{M}_{out} \rangle}{\partial \theta}\big|_{\theta}\Delta \theta$.
By matching the two equations we obtain 
$\Delta \theta =\frac{1}{\sqrt{p}}\frac{\Delta \hat{M}_{out} }{\big|\frac{\ud \langle \hat{M}_{out} \rangle}{\ud \theta}\big|}$. 
Let us notice that the error propagation analysis does not specify for which values of $p$ the estimate is unbiased.
\bibitem{nota2} In this Letter we will always consider the optimal condition 
$\cos(\theta_s-2\theta_c)=1$.
\end{thebibliography}
\end{document}